# Plasticity-induced restructuring of a nanocrystalline grain boundary network

Jason F. Panzarino[1], Zhiliang Pan[1], Timothy J. Rupert[1,2,*]

[1] Department of Mechanical and Aerospace Engineering, University of California, Irvine, CA 92697, USA

[2] Department of Chemical Engineering and Materials Science, University of California, Irvine, CA 92697, USA

[*]E-mail: **trupert@uci.edu**

**Abstract**

The grain boundary-mediated mechanisms that control plastic deformation of nanocrystalline metals should cause evolution of the grain boundary network, since they directly alter misorientation relationships between crystals. Unfortunately, current experimental techniques are unable to track such evolution, due to limits on both spatial and temporal resolution. In this work, molecular dynamics simulations are used to study grain boundary restructuring in nanocrystalline Al during both monotonic tension and cyclic loading. This task is enabled by the creation of new analysis tools for atomistic datasets that allow for a complete characterization and tracking of microstructural descriptors of the grain boundary network. Quantitative measurements of grain boundary character distribution, triple junction type, grain boundary plane normal, and other interfacial network characteristics are extracted and analyzed. The results presented here show that nanocrystalline plasticity leads to an increase in special boundary fraction and disruption of two-dimensional boundary connectivity, with the most dramatic evolution occurring in the smallest grain sizes.

# 1. Introduction

Nanocrystalline metals permanently deform through novel physical mechanisms [1, 2] that can be attributed to an increase in the grain boundary volume fraction. As average grain size ($d$) is reduced below ~10 nm, a crossover from intragranular dislocation-based plasticity to grain boundary-mediated plasticity occurs, with the mechanisms of grain boundary sliding [3, 4], grain rotation [5-8], and grain boundary migration [9] beginning to dominate. Because the grains themselves participate, these mechanisms are capable of altering not only grain shape but also interfacial structure during plasticity. For example, a rotating grain will change the misorientation relationship across its interfaces, potentially evolving grain boundary configurations and boundary energies [10, 11].

Atomistic models, most often molecular dynamics (MD) simulations, have played a crucial role in detailing nanocrystalline deformation physics and understanding how they alter microstructure. For example, Upmanyu et al. [12] used an atomistic model of an embedded nano-grain to show how simultaneous grain rotation and grain boundary migration reduces the overall interfacial energy of the system. Rotation rates and boundary migration velocities were highly dependent on the lattice misorientation and boundary energy, with larger misorientations and higher energies leading to relative increases in the rate of each mechanism. Similar work has also been used to uncover collective plastic rearrangement in polycrystalline systems. Hasnaoui et al. [13] used MD simulations of polycrystalline Ni ($d$ = 5 nm) subjected to constant tensile stress and observed an alignment of interfacial shear planes through collective grain rotation caused by intergranular slip. Rupert [14] also observed collective rotation of neighboring grains in simulated Ni ($d$ = 3 nm), reporting the development of shear localization pathways which extended across nanowire samples. A more quantitative approach was recently used by Panzarino et al. [15] to study relative contributions of each mechanism during mechanical cycling of polycrystalline Al ($d$ = 5 nm). These authors noted an overall reduction in grain boundary energy, which manifested as several occurrences of rotation-induced twinning as well as coalescence of low-angle grain boundaries within the microstructure. As a whole, prior work suggests that long-range evolution of the grain boundary network is likely during plasticity for extremely fine grain sizes.

Evolution of the grain boundary network would open the door for grain boundary engineering of nanocrystalline materials, where the fraction of boundaries with "special" properties can be altered with thermomechanical processing [16]. One approach for identifying special boundaries is to characterize interfaces using the coincident site lattice (CSL) model [17], which assigns a value ($\Sigma$) to each interface corresponding to the inverse of the maximum theoretical number of lattice sites which are shared between neighboring crystals, deeming those with $\Sigma \leq 29$ as special. A recent example of grain boundary engineering extended to nanocrystalline metals was presented by Bober et al. [18], who utilized transmission Kikuchi diffraction (TKD) inside of a scanning electron microscope to conclude that nanocrystalline Ni thin films ($d$ = 23 nm) subjected to thermomechanical cycling treatments will increase their $\Sigma 3$ boundary content as well as continually refine existing $\Sigma 3$ boundaries to a more perfect misorientation. This study confirmed the potential for altering grain boundaries through nanocrystalline deformation physics, but these authors focused on statistical boundary metrics rather than micro-mechanisms of evolution due to the inherent lack of temporal resolution provided by ex situ characterization. Kobler et al. [19] began to access time-resolved measurements by using automated crystal orientation mapping in the transmission electron microscope (ACOM-TEM) in combination with in situ straining. These authors found evidence of deformation-induced grain growth, grain rotation, and twinning/detwinning in nanocrystalline

Pd films with $d$ = 37 nm. However, because of the two-dimensional nature of data gathered by both TKD and ACOM-TEM, the orientation maps and boundary character distributions provided by such experimental techniques cannot confidently identify the grain boundary plane orientations (CSL only requires lattice misorientation between grains). In order to fully describe a grain boundary, the boundary surface normal (two parameters) in addition to the lattice misorientation relationship between grains (three parameters) must be determined [20].

Fortunately, since MD simulations can provide the exact positions of atoms during the entire course of deformation, all of the required information is available to extract and observe the intricacies of grain boundary network restructuring as it is occurring, on a femtosecond timescale. Currently, such data mining is still a daunting task, even with the advent of several tools and metrics capable of quantifying nanoscale microstructural features. Several researchers have constructed analysis methods which can identify grains [21-23], interface atom types [23], and even extract entire dislocation networks including boundary dislocations [24], but there are currently no tools that can extract and track grain boundary character from raw atomistic simulation data. In order to truly understand how grain boundary mediated plasticity can evolve grain boundary networks, new algorithms must be developed which are capable of segmenting interfacial regions and then extracting features like grain boundary surface normals, grain orientation/disorientation relationships, and triple junction structure, while simultaneously tracking these features as they evolve throughout a simulation. Such analysis tools would provide fully characterized five parameter grain boundary data with spatial and temporal resolution that is currently inaccessible with experimental techniques.

In order to understand how plasticity at the nanoscale evolves grain boundary networks, we use MD simulations to subject nanocrystalline Al to both monotonic tension as well as mechanical cycling. Two grain sizes, 5 nm and 10 nm, were examined so that differences in grain boundary network evolution associated with different plasticity mechanisms can be observed. In addition, the effect of temperature on boundary evolution during plasticity was also studied. In order to provide a truly quantitative analysis of the grain boundary network, we have developed several new techniques for characterizing and tracking grain boundary features and incorporated these tools into a recently developed analysis algorithm for atomistic datasets [21]. The tool developed here is able to identify all grain boundaries, triple junctions, and vertex points while also providing interface character and network connectivity information using methods familiar to the experimental community. We find that special boundary fraction dramatically increases with deformation for $d$ = 5 nm, with elevated temperature and repetitive loading giving the largest changes. This evolution mainly manifests as an increase in the fraction of $\Sigma 3$ and $\Sigma 11$ boundaries, which we show is related to restructuring to find low-energy boundary configurations. This evolution is contrasted with boundary network evolution during thermal annealing, to highlight differences that are characteristic of mechanically-driven structural reorganization.

## 2. Simulation Methods

Samples with average grain sizes of 5 nm and 10 nm were constructed using the Voronoi method with random Euler angles assigned to each grain nucleation site [4, 25]. In addition, a minimum separation distance between grain nucleation sites was employed which allows for more equiaxed grains and a tight grain size distribution to more efficiently analyze the response of each specific grain size. For each sample, the same set of Euler angles was used to ensure that the starting texture and grain boundary structure would be identical despite the difference in grain size. The Large-scale Atomic/Molecular Massively Parallel Simulator (LAMMPS) [26] was used and all atomic interactions were described by an embedded atom method (EAM) potential for Al developed by Mishin et al. [27]. This many-body potential was developed using a combination of *ab initio* and experimental data with the intention of simulating internal defects and plasticity in Aluminum and was found to accurately describe point defects, planar faults, grain boundaries, and experimental values of stacking fault energy. A 2 fs time step was used for time integration during the simulation with periodic boundary conditions applied in all directions. Any overlapping atoms which were separated by less than 2 atomic radii were removed and the final structures were then relaxed using a conjugate gradient minimization with an energy tolerance of $10^{-6}$ eV and a force tolerance of $10^{-6}$ eV/Å. The resulting structures were fully dense and free of stored dislocations, containing 48 grains. The $d = 5$ nm sample contained 180,982 atoms with a cubic simulation cell side length of 146.5 Å and the $d = 10$ nm sample contained 1,480,503 atoms with a side length of 294.4 Å. The samples were then annealed at 600 K for 100 ps to remove excess grain boundary dislocations and free volume [28-30]. This step ensures that any observed structural evolution is not merely a byproduct of unstable high-energy interfaces caused by the construction technique used to create the samples. After equilibration, samples were cooled at a rate of 30 K/ps until the desired testing temperature was reached. Although these Voronoi samples are the main focus of this work, we also compare with nanocrystalline atomistic models created using two other construction techniques that give slightly different starting microstructures. Further details of this additional analysis is provided in Section 4.3.

For mechanical testing, monotonic tension to 10% true strain and mechanical cycling were performed. Cycling was achieved through tension load-unload cycles. Specimens that had been loaded past the yield point to 5% true strain were first unloaded to 3% strain and then pulled back to 5% strain, with each unloading-loading pair representing a single cycle. This process of mechanical cycling was repeated for a total of 10 cycles. Both sets of mechanical testing simulations were run at temperatures of 300 K, 450 K, and 600 K, using a true strain rate of $5 \times 10^8$ s$^{-1}$ while employing a Nose-Hoover isothermal-isobaric ensemble in order to keep zero stress in the lateral sample directions. A recent study by Zhang et al. [31] showed that, for nanocrystalline Cu, the grain boundary mediated modes of plastic deformation are not strain rate dependent in the range of $1 \times 10^7$ s$^{-1}$ - $1 \times 10^{10}$ s$^{-1}$. Figure 1 illustrates the deformation methodology and typical results for monotonic tension (black curve) and mechanical cycling (red curve) of the $d = 5$ nm sample at 300 K. For monotonic tension, atomic snapshots were output at the starting configuration and at each strain increment of 1%. For the tension unload-load cycling, simulation snapshots were recorded before deformation (0% strain) and each time the sample reached 5% strain. These snapshots were subsequently analyzed using the Grain Tracking Algorithm (GTA) [21], after incorporating the improvements detailed in Section 3, and visualization of the resulting data was performed by the open source particle visualization tool OVITO [32]. The updated GTA is freely available through the corresponding author's research website or can be obtained by contacting the corresponding author.

## 3. New Analysis Methods

The GTA was recently developed to provide grain identification, texture analysis, and grain structure evolution for atomistic simulation outputs. As outlined in a previous article that summarizes the initial development of the tool [21], spatial coordinates of the atoms along with a crystallinity description (e.g., centrosymmetry parameter (CSP) [33] or common neighbor analysis (CNA) [34]) are all that is required to identify and track crystallites and their orientations during the simulation. CSP is a measure of the local lattice disorder surrounding an atom and was chosen here to distinguish between crystalline and non-crystalline/defect atoms within our samples. A threshold of CSP $\geq 2.83$ Å$^2$ was chosen to signify non-crystalline atoms in this study based on the Lindemann-Gilvarry rule [35], which establishes the maximum bond length allowable for an atom contained in a crystalline environment before melting occurs. Once atoms have been separated into crystalline and non-crystalline, the GTA calculates the lattice orientation at each crystalline position within the sample and uses this information to segment the sample into individual grains. The grain segmentation technique compares the disorientation angle (smallest symmetrically equivalent misorientation angle) between lattice orientations measured at each atomic position and compares these disorientations with an allowable cut-off angle. Any neighboring atoms which fall within this cut-off are deemed part of the same crystal. To improve grain segmentation during high temperature, all nearest neighbor vector combinations which determine a set of axis at each crystalline point are averaged to allow for more robust orientation determination. Using fcc as an example, the orientation determination procedure outlined by Panzarino et al. [21] is repeated for all 12 nearest neighbor position vectors providing 12 sets of axis which are then averaged. In addition to grain segmentation, the GTA algorithm also provides a mapping between time steps, allowing for the tracking of features such as grain size, grain rotation, and grain sliding. Figure 2(a) shows the $d = 10$ nm microstructure with atoms colored according to grain number, as well as an accompanying pole figure indicating a randomly oriented starting texture of the grains.

To allow for quantitative analysis and tracking of important features of the grain boundary networks, the GTA was updated in several important ways. Having calculated grain structure and orientation information, the algorithm now proceeds to distinguish between the several types of interfacial atoms within the microstructure. Boundary atoms which reside between two grain neighbors are deemed grain boundary plane atoms, while those with three neighbors belong to triple junctions and four or more neighbors are classified as vertex point atoms. Remaining non-crystalline atoms which have only one grain neighbor (meaning they reside within the grains themselves) are deemed intragranular defect atoms can be associated with a dislocation, stacking fault, vacancy, or interstitial. An example of this classification can be seen in Figure 2(b), where all crystalline atoms have been removed. Intragranular defect, grain boundary plane, triple junction, and vertex atoms are colored light blue, green, yellow, and red, respectively. Examples of a stacking fault and a vacancy are labeled. Figure 2(c) displays an example of an isolated grain extracted from the computational model using the same atom classification scheme. A similar boundary atom classification was used by Xu and Li [23], but their approach obtained grain numbers and crystal orientations from the initial Voronoi construction. An outward layering method was then used to identify the different boundary atom types. Our method is similar in its classification scheme but also allows for the indexing of microstructures with no a priori knowledge of grain locations or orientations. Such a feature is essential for tracking boundary features during a simulation, as the material evolves.

After indexing atom types, the updated GTA then calculates the disorientation angle and axis of rotation between all neighboring grain pairs so that features such as $\Sigma$ type in the CSL framework can be calculated. It is important to note that the CSL framework is simply one choice for describing grain boundary character distributions and there are others such as grain boundary plane distribution [36] and disorientation axis distribution [37]. Since the GTA calculates all five degrees of freedom for a boundary, any of these methods is an option. We choose to use $\Sigma$ type here to describe boundary character so that our findings can be readily compared to experimental reports. The Brandon criteria [38, 39] defines the maximum allowable mis-misorientation offset,

Δθ, of an interface from a perfect Σ misorientation relationship and is typically defined as $\Delta\theta_{max} = 15°/\Sigma^{1/2}$. Here we use the Brandon criteria, but again a simple modification the GTA script would allow for the use of other metrics. Figure 3(a) shows a nanocrystalline sample with all grain boundary plane atoms colored according to their disorientation angle. The lighter colors indicate higher angle grain boundaries and the boundaries highlighted in blue are special boundaries, or interfaces with Σ ≤ 29 in the CSL notation.

In order to fully characterize a grain boundary and its structure, the grain boundary normal must also be determined. Since the CSL model is only related to the angle and axis which describe misorientation, each Σ type can actually exhibit infinitely many variations depending on the dividing plane which separates the two crystals. A recent study by Homer et al. [40] illustrates the importance of boundary plane-property relationships in Ni and Al bicrystals using a technique developed by Patala and Schuh [41] to rewrite grain boundary normals in terms of a symmetry-capturing fundamental zone. Their results showed that for various special boundaries, grain boundary energy and excess free volume can vary dramatically depending on the grain boundary plane orientation, with only specific boundary normals exhibiting energy minimums. This additional boundary data is essential for understanding the role of interface planes in grain boundary network evolution and several other emerging studies also highlight the importance of a complete grain boundary character analysis which includes such information [42-44].

To obtain the grain boundary normal, the updated GTA segments the grain boundary plane atoms into planar sections based on whether or not each atom and its neighbors constitute a relatively flat section, which allows for the identification of faceted boundaries. First, local surface normals are calculated at each grain boundary atom using a singular value decomposition to find a best-fit plane for each atom and a surrounding cluster of its nearest neighbors. Singular value decomposition determines a plane which minimizes the square sum of the orthogonal distances between the points and the best-fit plane. For nanocrystalline microstructures, a cluster of 50 grain boundary neighbor atoms appropriately captures accurate interface normals at each atom, with an example shown in Figure 3(b). This process is repeated for all of the atoms in the boundary and a normal is assigned to every atomic position. When building up a grain boundary section, a grain boundary normal variation angle, $\theta_n$, must be selected which provides a maximum allowable normal variation between neighboring atoms. Figure 3(b) illustrates this concept by showing three neighboring atoms with their normal vectors assigned and how their orientations can vary by $\theta_n$. A convergence study was performed to find an adequate grain boundary cutoff angle. The grain boundary network of the 10 nm average grain size sample ($T = 450$ K) was analyzed at 10% strain while varying $\theta_n$ from $4-12°$ and recording the amount of grain boundary atoms which were unable to be allocated to grain boundary sections. The results of the convergence study can be seen in Figure 3(c). A value of 8° allows for reasonable accuracy and the ability to segment flat sections of grain boundary, while also ensuring that there are only a small number of non-allocated interface atoms. For stricter cutoff angles, many atoms within grain boundary planes would be unallocated. Alternatively, larger cutoff angles would not adequately segment curved interfaces into planar sections. Proper selection of this normal cutoff angle is important for nanocrystalline systems with very small grain sizes, where planar grain boundary sections tend to be very small and grain boundaries can be highly curved. With boundary plane normal identified, all five grain boundary parameters in an atomistic sample can be fully characterized.

With boundaries completely described, higher level topological features such as triple junctions can also be quantified. Depending upon how many of the connecting interfaces exhibit special character, triple junctions can be assigned a designation number or triple junction type.

Since connectivity of high energy interfaces also has implications regarding percolation of fracture-susceptible boundaries [45, 46], several past studies have focused efforts on characterizing triple junction type distribution and cluster connectivity using two-dimensional computational models or orientation imaging microscopy scans [47-50], with a three-dimensional numerical study utilizing a tetrakaidecahedra model for the grain shapes [51]. In order to be consistent with this previous work, triple junction types are assigned here as follows: junctions with no attached special boundaries are deemed type 0, junctions with one special boundary are called type 1, and so on. Figure 3(d) shows the triple junction network in the $d = 10$ nm sample where the atoms are color coded by type number. Analyzing the density as well as connectivity of these special interface types can help illuminate the role of triple junction topology during microstructural evolution.

## 4. Microstructural Evolution During Plastic Deformation

*4.1. Monotonic Loading*

First, the effect of simple monotonic plasticity on the grain boundary network is studied. During deformation, evolution of the grain boundary character distribution was analyzed and special boundary fractions are reported. The fraction of boundary atoms which are detected as having $\Sigma \leq 29$ is plotted as a fraction of the total number of grain boundary plane atoms. Since this metric is an atomic fraction, our calculation will be similar to length fraction, but with an added grain boundary thickness component which we expect to have minimal effect. Grain boundary special fractions are presented in Figure 4(a) and 4(b), as a function of applied strain, for the $d = 5$ nm and $d = 10$ nm samples, respectively. As the 5 nm grain size samples are deformed, the fraction of special boundaries shows some fluctuations, with the general trend that the special fraction increases during the majority of the plastic deformation range. These increases in special fraction occurred through both an increase in the length of existing special boundaries as well as the development of new special boundaries that were not present before deformation. Occasional jumps in the data are a result of large boundaries which evolve into or out of special character as additional grain rotation occurs, so it is most instructive to look at general trends. Most obvious is the fact that, tension at higher temperatures led to faster increases to the special boundary fraction. Further inspection showed that boundaries with low potential energy were more likely to cease rotation and hold a consistent disorientation relationship after their formation. For example, $\Sigma 3$ boundaries became sessile if {111} boundary planes were formed. However, other $\Sigma 3$ boundaries with randomly oriented boundary planes were not guaranteed to keep their character through the rest of the deformation process and occasionally moved away from a special configuration. Similar behavior was observed for other special boundaries as well. This importance of grain boundary plane on boundary stability highlights the power of being able to extract such information from atomistic datasets, a capability enabled by the updated GTA.

For the 10 nm samples, the grain boundary special fraction remained relatively constant with some fluctuations around the starting value. This difference can be attributed to the larger grain size, where dislocation-based mechanisms start to dominate. Schiotz and coworkers [52, 53] noted an inverse Hall-Petch relationship in the flow stress of simulated Cu when average grain size fell below 10 nm and further inspection showed that plastic strain was highly concentrated at the grain boundaries in the finest grained samples. Kadau et al. [54] also showed that a transition in plasticity mechanisms occurs near this grain size for Voronoi-created Al samples tested by MD. In our analysis, grain rotation is found to be significantly restricted for the larger grain size and an increase in the density of residual stacking faults left in the microstructure was observed for the 10 nm sample, which are both consistent with a shift toward dislocation-based plasticity mechanisms. In addition, no significant reduction in the potential energy of special boundaries was measured during deformation of the larger grain size sample.

The accompanying triple junction network was also tracked, with Figure 4(c) and (d) showing the evolution of type 0 triple junctions as a function of applied strain. The measurement of triple junction fraction was carried out in a fashion similar to special boundary fraction, with the total number of atoms residing in each triple junction type divided by the total number of triple junction atoms present. The decrease in type 0 triple junction fraction for the $d = 5$ nm sample is a direct consequence of the increase in special boundary fraction with applied plastic strain. With more and more special boundaries, it becomes less likely that a triple junction will not have a special boundary attached to it. Because of the limited change in special fraction for the $d = 10$ nm sample, there was no clear trend in type 0 evolution. The remaining triple junction types were also analyzed, but no obvious trends were found and no additional insight was obtained through the metric that could not otherwise be portrayed in the special fraction data.

Connectivity and topology of the boundary network was also analyzed to understand if any long-range evolution is occurring. Since it is known that connectivity of the grain boundary network relates to intergranular failure such as fracture and corrosion [46, 55-59], the size of connected "random boundary clusters" of non-special boundary atoms was measured. Figure 5 compares a two-dimensional slice of the $d = 5$ nm sample at strains of 1% and 10% at 600 K. Special boundaries, triple junction types greater than zero, and vertex points which adjoin these special features were removed from the slices, since special interfaces tend to be less susceptible to intergranular failure. The remaining grain boundary atoms are segmented into clusters and colored using the cluster analysis tool developed by Stukowski [32] which groups clusters of particles using a user-specified cutoff radius, with a cluster radius of 6 Å used for Figure 5. A comparison of Figure 5(a) and (b) shows that there is significant break-up in the random boundary cluster size with increased plastic strain. This means that, for example, a crack moving from left to right through the microstructure would not be able to find a path which only propagates along random, non-special boundaries. The connectivity of random boundary clusters in two-dimensions is commonly reported in the experimental literature for grain boundary engineering (see, e.g., [16, 60]). However, extension to three-dimensional analysis, available in our simulation method but not generally available from experiments, shows no large changes in cluster number or average size. Even though the network is broken up locally along a two-dimensional slice, the additional dimensionality allows for more percolation pathways and any breakups can be potentially bypassed through another route present in the grain boundary network. In fact, Figure 5(c) is the same sample at 10% strain with blue atoms illustrating the single three dimensional cluster that exists despite the fact that the special boundaries (green), triple junctions of type greater than zero (yellow), and vertex points (red) were removed from the cluster analysis. Again, this observation highlights the utility of atomistic modeling for analysis of boundary network features with additional detail.

*4.2. Cyclic Loading*

According to Fig. 4(a), the majority of special fraction increase occurred within the intermediate 3-8% strain range and higher temperatures magnify this effect. In an attempt to drive further reconstruction of the grain boundary network, mechanical cycling in the range of 3-5% true strain was simulated. Figures 6(a) and (b) plot the special boundary atomic fractions, with the X-axis beginning at the starting configuration of 0% strain (labelled "Start") and then showing data from the end of each cycle (Cycle 0 is the initial loading to 5% strain). It is clear from Figure 6(a) that cycling at higher temperature gives a faster increase in special boundary fraction for the $d = 5$ nm sample. In fact, cycling at 300 K does not clearly alter the special fraction in an obvious manner. Like monotonic loading, the observed evolution is due to both an increase in the length of existing special boundaries as well as the emergence of new special boundaries. Figure 6(b) shows the special boundary fraction evolution for the $d = 10$ nm grain size as a function of cycle number. There is a noticeable decrease in the special boundary fraction during the 600 K cycling, as well as a small decrease for cycling at 300 K and 450 K. Generally, there is significantly less evolution for the larger grain size. For the 5 nm grain size, it is instructive to compare the special boundary fraction achieved through mechanical cycling to that obtained with monotonic tension. By the 10th cycle, the $d = 5$ nm sample tested at 600 K contains a 0.175 fraction of special boundaries (or an increase of ~75% from the starting value) whereas simple tension only resulted in a fraction of 0.12 or less for all strains at or below 5%. Even considering all possible strains up to 10%, monotonic tension only results in a 0.14 special boundary fraction. While monotonic tension and mechanical cycling both increase the special boundary fraction in these samples, cyclic loading allows for extra boundary rearrangement to occur.

The main difference in grain boundary network evolution between these two grain sizes can be attributed to increased activation of grain rotation in the smaller grain size sample, as shown in Figure 6(c) where average grain rotation is plotted as a function of cycle number for the 600 K simulations. Here, grain rotation is measured as the angular disorientation of a grain from its starting configuration (Start), before any deformation was imposed. As a whole, the $d = 5$ nm sample experienced roughly three times as much grain rotation as the 10 nm grain size sample. It is also important to note that these values are averaged over all grains in the sample. For example, there were several grains in the $d = 5$ nm sample which rotated more than 5 degrees, but none in the $d = 10$ nm sample. The error bars in Figure 6(c) show the standard deviation of the grain rotation angles. This grain rotation was often accompanied by simultaneous grain boundary migration of known high-energy interfaces. For example, non-coherent sections of $\Sigma 3$ boundaries migrated to allow for lengthening of coherent $\Sigma 3$ sections. This type of evolution was found to occur for other types of special boundaries as well. Figure 7 displays a more detailed picture of the resulting evolution for all special boundaries in the $d = 5$ nm and $d = 10$ nm samples that were mechanically cycled at 600 K. $\Sigma 13$, $\Sigma 17$, $\Sigma 19$, $\Sigma 21$, $\Sigma 25$, $\Sigma 27$, and $\Sigma 29$ boundaries each have "a" and "b" variations that were tabulated but these are reported here in a combined fashion using their respective $\Sigma$ value. Variations in CSL types are a result of specific $\Sigma$ boundaries which have multiple disorientation angle/axis pairs which result in the same reciprocal coincident site density [61]. Figure 7(a) indicates a distinct increase in $\Sigma 3$ and $\Sigma 11$ boundaries, while other $\Sigma$ types show no clear trend or even slight decreases (see the $\Sigma 15$ population). In contrast, Figure 7(b) illustrates noticeably lower special fraction evolution within the $d = 10$ nm microstructure during cycling and lack of a clear trend.

It is widely reported that special boundaries can exhibit low energy, but this is really only rigorously true for certain interface plane normal values [11, 62]. Additionally, several computational works [12, 15, 63] and experimental studies [64, 65] provide evidence that grain rotation is a mechanism by which interfacial energy can be reduced. Therefore, one explanation for the increase in $\Sigma 3$ and $\Sigma 11$ boundaries is that rotation-mediated boundary rearrangement allows for the emergence of these special boundary types once favorable grain boundary planes are achieved. Early experimental works studied rotation-induced formation of special boundaries through the use of single crystal particles sintered to flat plates [65, 66]. Since the particles were sitting freely on top of the substrate, they were unconstrained by any neighboring crystals and thus free to rotate during sintering. Herrmann et al. [65] discovered the strong emergence of Σ3 as well as Σ11 boundaries during the sintering of both Cu and Ag particles in this type of experiment. Because our present study shows high levels of grain rotation for $d = 5$ nm, this suggests that nano-grains are able to rotate to low-energy cusps in the boundary energy landscape in a similar fashion.

Analysis of several specific special boundary types is presented in Figure 8(a), where the average atomic potential energy is plotted versus cycle number. The remaining special boundary types that are not shown follow roughly the same trend as the Σ5 boundaries. The average potential energies of crystalline atoms as well as random boundary atoms are also presented as grey and black data points, respectively. Even before loading of any kind, the pre-existing Σ3 and Σ11 boundary atoms have lower energy than other grain boundary atoms, foreshadowing the restructuring to come that will increase their fraction within the microstructure. The Σ3 boundary atoms approach a lower energy state with added cycles, especially after the initial pull and first cycle. Additional analysis of the Σ3 boundaries shows that the average mis-misorientation, or angular deviation from the perfect CSL disorientation, decreases with cycling, meaning that they are becoming closer to a perfect Σ3 misorientation. This mis-misorientation evolution of Σ3 boundaries for the $d = 5$ nm samples is shown in Figure 8(b). The majority of the Σ3 boundary planes evolve to coherent {111} planes by the 10$^{th}$ cycle, while those that did not were the facets that enabled {111} plane formation elsewhere. Visual evidence of this is shown in Figure 8(c), which displays all Σ3 boundaries at Cycle 10 for the $d = 5$ nm, 600 K sample. In contrast, the average energy of all Σ11 boundaries was relatively low in the original configuration and only fluctuated slightly with cycling. The Σ5, Σ9, Σ27 boundaries did not undergo significant evolution, with their energies roughly matching the average atomic energy of the remaining random boundaries. Data for the Σ9 and Σ27 boundaries does not appear across all cycles because they were not always present in the sample. Finally there is even some slight reduction in the average energy of crystalline atoms which can probably be attributed to structural relaxation of the system during early loading.

Knowing that Σ3 and Σ11 boundaries were evolving to reduce interfacial energy, site specific occurrences of the formation of these boundary types were then studied to obtain more insight into the underlying restructuring mechanisms. As an example, Figure 9(a)-(c) follows the structural rearrangement and formation of four special boundaries which all border a common grain. For clarity, the crystalline atoms as well as triple junction and vertex point atoms have been removed. By Cycle 4, three of the interfaces have become special, having evolved from previously random boundary types. The details of each misorientation relationship are listed below each snapshot and color coded according to the respective boundary. After first listing the Σ value, the disorientation angle between grains and the mis-misorientation angle are provided. Next, the rotation axis is listed, followed by the grain boundary normal vector that is written in terms of the crystal orientations of the two neighboring grains. For those boundaries which are highly curved, the normal vectors are not listed until an adequately planar interface or faceted structure develops. Continued cycling results in the formation of a Σ27b boundary and additional structural rearrangement of the interfaces. First, the Σ3 boundary reduces its mis-misorientation from 5.5° to 1.89° by the last cycle. Cycling also allowed for an increase in the surface area of the Σ3 interface. The Σ25b, which did not have an ordered boundary plane, shrinks during cycling. The Σ27b forms at the end of cycle 7, but there is little change in mis-misorientation with continued cycling. The most interesting evolution is observed for the Σ11 boundary. This boundary begins to facet during cycling, to reduce the boundary energy at the cost of increased surface area. Facets are denoted in Figure 9(b) and (c) using dashed lines. Further inspection of the grain boundary energy and facet structure evolution is shown in Figure 9(d)-(f), which is a slightly rotated view of the Σ11 boundary colored by potential energy. Three facets at the end of Cycle 8 are denoted by black arrows and have low relative potential energy, especially at the innermost atomic layer. The facet steps continue to restructure during the next two cycles, resulting in additional length of the low energy segments. Figure 9(g) shows the detection of the facet planes as computed by the GTA with character information listed below the image. These low energy planes are Σ11 {113} grain boundary sections, which are known to be a minimum energy cusp for symmetric tilt boundaries [40].

*4.3 Effect of Starting Configuration*

As some recent studies have questioned whether a Voronoi grain structure is a realistic representation of real polycrystalline microstructure [67-71], it is nature to ask whether the results reported above are general findings or related specifically to the evolution of a Voronoi grain structure. To rule out any artifacts due to any specific sample generation technique, our original Voronoi samples were compared to two additional samples generated from separate three-dimensional isotropic grain growth models developed by Lazar et al. [72] and Syha and Weygand [73], with a focus on cycling of the $d = 5$ nm structures at 600 K since this was the condition that showed the most evolution. These models allow for different, potentially more realistic, distributions of grain size as well as grain boundary topological features and boundary curvatures that may not be present within an as-assembled Voronoi construction. The Front-Tracking implementation of Lazar et al. [72] is known to reproduce interesting geometrical features such as two-sided faces and three-faced bodies, which occur as the transitioning microstructure coarsens to a steady state. The Vertex Dynamics model of Syha and Weygand [73] includes an additional force term which incorporates the effects of inclination and misorientation dependent grain boundary energy. Both models satisfy the MacPherson-Srolovitz relation for grain growth rates in three dimensions [74]. Lazar et al. [72] and Syha and Weygand [73] developed coarsened microstructures by starting with grains that were initially generated using the Voronoi construction and then grown using their Front-Tracking and Vertex Dynamics algorithms, respectively. In the present study, we have taken the grain and boundary locations from these two studies and filled them with atoms corresponding to randomly oriented crystals using the open source code NanoSCULPT developed by Prakash et al. [75]. The overall sample size was scaled to microstructures that are comparable to our as-assembled $d = 5$ nm grain size, before then subjecting both of these samples to the 600 K anneal for 100 ps. These two samples were then cycled at 600 K in order to observe and compare the resulting grain boundary network evolution. For the remainder of this section our three different starting configurations will be referred to as Voronoi, Vertex Dynamics, and Front-Tracking to facilitate comparison between the different sample generation methods.

Visual comparisons of all three starting configurations can be seen in Figure 10(a)–(c) which are sliced along a (111) sample plane and color coded according to grain identification by the GTA. The Vertex Dynamics and Front-Tracking samples contain larger numbers of grains, which allows for improved evolution statistics as well as some insight into whether our previous results are affected by the selected sample size. The Vertex Dynamics sample contained 573,690 atoms with a cubic simulation cell measuring 146.5 Å in length and the Front-Tracking sample contained 632,469 atoms with a cubic simulation cell measuring 222.0 Å. The grain size distribution for all three samples is presented in Figure 10(d), where there is a clear contrast between the sharp distribution in the Voronoi sample as compared to the broader distributions found within the grain growth models. The exact average grain sizes for all three models after the 600 K anneal were computed to be 5.4 nm, 6.7 nm, and 6.4 nm for the Voronoi, Vertex Dynamics, and Front-Tracking samples, respectively, based on an equivalent spherical diameter calculated from the atomic volume of each GTA identified grain.

Special boundary and $\Sigma 3$ fraction evolution is shown in Figure 11(a) for all three sample generation methods as a function of cycle number. Even with a varying initial distribution of special boundary content there is a consistent increase in special boundary fraction during cycling for all $\Sigma \leq 29$. The $\Sigma 3$ content evolves slightly faster in the Voronoi samples, which could in part be attributed to the slightly smaller average grain size and an increased contribution from grain rotation. The slight reduction in $\Sigma 3$ evolution may be connected to the fact that the wider grain size distribution of the Vertex Dynamics and Front Tracking contain a few grains in the 10 nm range, where we previously reported less $\Sigma$ fraction evolution in Figure 7(b). In the end though, significant evolution of both special boundary and $\Sigma 3$ fractions are observed with cycling, demonstrating that the observations made in this paper are applicable to a general nanocrystalline grain structure. The complete evolution of each special boundary type is presented in Figure 11(b) and (c) for the Vertex Dynamics and Front-Tracking, respectively. An interesting trend across the two new samples is the increase in $\Sigma 3$ and $\Sigma 11$ boundary fractions, similar to the Voronoi observations. The three models produce differing initial $\Sigma$ fractions, meaning that some details of the grain boundary character evolution are different. For example, the Front-Tracking sample has the largest fraction of $\Sigma 5$ boundaries. Since these are relatively high energy boundaries and the system evolves to a lower energy state during cycling, the $\Sigma 5$ fraction rapidly decreases during cycling. As another example, the Vertex Dynamics sample has an increasing Σ13 fraction, while maintaining the relatively large initial fraction of Σ29 and Σ21 boundaries. As a whole though, the comparison of the three different samples allows general conclusions to be drawn. Mechanical cycling at elevated temperature allows the grain structure to rearrange, with an increase in low energy Σ3 and Σ11 fractions in all cases.

*4.4. Comparison with evolution during annealing*

The results shown in the prior two sections clearly show that plasticity is capable of restructuring nanocrystalline grain boundary networks so that lower energy configurations can be found. Thermal annealing is another common way to drive microstructural evolution toward a lower energy state, and can serve as a comparison point for the mechanically-induced restructuring already described. To facilitate such a comparison, the $d$ = 5 and 10 nm samples were also subjected to thermal treatments at 800 K. The 600 K equilibration samples were heated in a linear fashion to 800 K over the course of 100 ps, then held at this temperature for 1 ns with atomistic snapshots analyzed in 100 ps time intervals. The grain boundary special fractions during this annealing are presented in Figure 12(a). There was a dramatic increase in special boundary fraction for the 5 nm grain size, but no increase for the $d$ = 10 nm sample. An in-depth analysis of boundary evolution in the 5 nm sample is shown in Figure 12(b). Most noticeable is the rapid increase in Σ3 boundary fraction during annealing. After reaching 800 K (0 ps), partial dislocation emission along successive {111} planes allows for pre-existing stacking faults to form into small annealing twins. These twinned regions remain fixed as the remaining grain boundaries within the sample migrate rapidly in directions parallel to the {111} planes, to lengthen the coherent twin interfaces. The majority of the Σ3 content was observed to develop in this manner with minor evolution occurring to the preexisting non-coherent Σ3 boundaries, the majority of which disappeared from the microstructure soon after heating. Bringa et al. [76] discovered five-fold twinning in simulated Cu ($d$ = 5 nm) annealed at 800 K, explaining that high local stresses within the grain boundaries combined with elevated temperature can allow for emission of twinning partials during annealing. Although lengthening of Σ3 boundaries dominates network evolution during annealing, some small increase to the Σ5, Σ7, and Σ11 population also occurs. Unlike the mechanical loading case where Σ11 boundaries are the second most frequent special boundary, Σ5 content increases most quickly during annealing. However, after ~500 ps the average grain size becomes very coarse and several grains approach the length of our simulation cell.

In order to show the typical physics of network restructuring during annealing, a representative example of Σ3 boundary lengthening is shown in Figure 13(a)-(c). As the annealing simulation begins, the highly curved, random boundaries start to migrate toward their centers of curvature (black arrows). This migration allows a coherent (111) interface to increase its length dramatically and obvious grain growth has occurred by 200 ps. Further inspection shows that this is not the only special boundary which formed along this grain pair during the annealing process. Figure 13(d) presents a side view showing the formation of a second twin boundary and a Σ7 boundary. A facet step present in Twin 1 is also observed. Additional annealing allowed for the continued migration of adjacent random boundary segments, which lengthens the Σ3 boundaries and Σ7 boundary until they meet and lock in position at 500 ps, as shown in Figure 13(e). The Σ7 boundary obtains a distinct (111) orientation normal by this time, which is a minimum energy configuration for the Σ7 boundary type [40]. Network rearrangement during annealing therefore leads to different special boundaries (Σ5 and Σ7 are common) and different mechanisms of evolution (curvature-driven grain growth).

**Conclusions**

In this work, a quantitative analysis of the grain boundary network and its evolution during monotonic and cyclic plastic deformation was presented. Since experimental techniques lack the temporal and spatial resolution for such measurements, molecular dynamics were used to simulate deformation while new analytical tools are created to quantify important grain boundary features. By investigating two grain sizes that span the Hall-Petch breakdown, we find that finer nanocrystalline grain sizes experience more evolution of their grain boundary network during plasticity. In addition, the magnitude of network restructuring is highly temperature dependent. Several important conclusions can be drawn from this work:

- Analysis tools were developed that allow for quantification of the grain boundary network in atomistic models. The five degrees of freedom associated with any grain boundary section can be measured, as well as features associated with triple junctions and vertex points. With this information, grain boundary character distributions and network topology/connectivity can be characterized with nanometer and femtosecond resolution.
- Both monotonic and cyclic plasticity drive an increase in the special boundary fraction for $d = 5$ nm. The majority of the increase was associated with higher Σ3 and Σ11 boundary fractions, through both the lengthening of existing boundaries and the creation of new boundary sections. For the Σ3 boundaries specifically, mis-misorientation decreases during cycling and low-energy (111) boundary planes are often found. Similar faceting to create low-energy Σ11 interfaces is also observed. The preference for these low-energy grain boundary types suggests that the grain structure is rearranging to reduce system energy and find a more stable configuration.

- Negligible restructuring was found for the $d = 10$ nm sample. This lack of evolution can be linked to the reduced activation of collective plastic mechanisms such as grain rotation and the increase in dislocation activity.

- Even when varying the method used to create the starting microstructure, which also alters the initial distribution of special boundary content, a consistent increase in $\Sigma \leq 29$ fraction is found for small grain sizes during mechanical cycling at elevated temperature. The formation of low energy Σ3 and Σ11 boundaries is common to all samples.

- Annealing also leads to a reduction in system energy, but there are important differences in the special boundary types which form and in the mechanisms for formation. Σ3 boundaries lengthen by curvature-driven grain growth, where fast migration of random boundaries parallel to the (111) twin planes drastically increases Σ3 content.

This study shows that the collective deformation physics associated with nanocrystalline metals can lead to evolution and restructuring of the grain boundary network. Grain boundary networks are therefore very dynamic at the finest nanoscale grain sizes and grain boundary engineering through new mechanisms is an intriguing possibility for these materials.

**Acknowledgements**

We gratefully acknowledge support from the National Science Foundation through a CAREER Award No. DMR-1255305. The authors also thank Dr. Daniel Weygand for supplying the

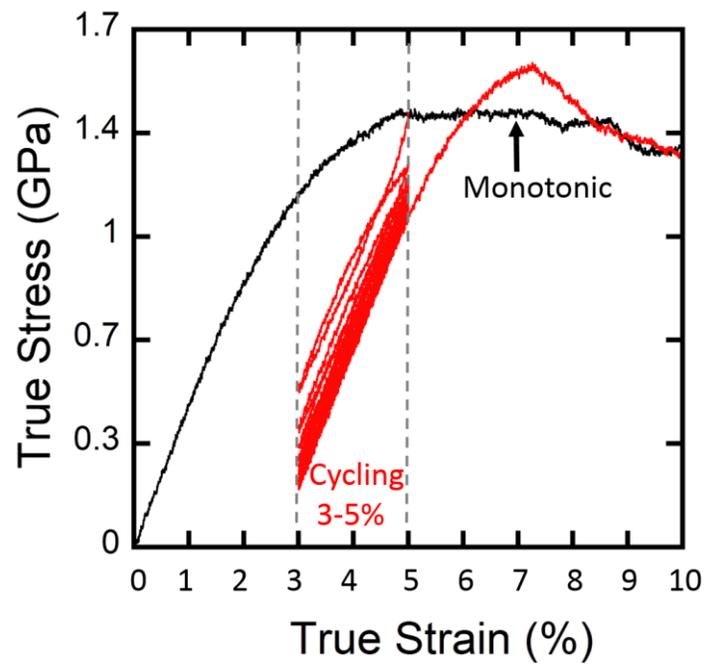

**Figure 1.** Mechanical response of *d* = 5 nm sample loaded monotonically (black line) and cyclically (red line) at 300 K. Dotted grey lines illustrate the mechanical cycling procedure. Identical loading procedures were performed at 450 K and 600 K.

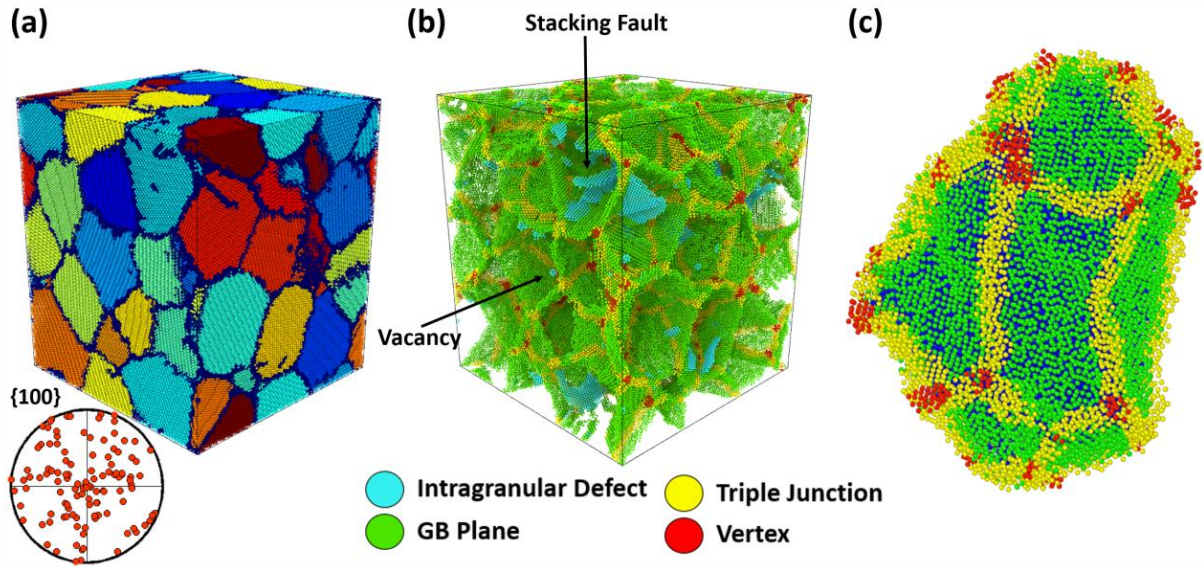

**Figure 2.** (a) The *d* = 10 nm sample at 10% true strain and 600 K with all grain identified. The {100} pole figure indicates random texture throughout the sample, even after tensile deformation. (b) The grain boundary and intragranular defect network of the same sample with atoms colored according to defect type. Blue atoms are intragranular defects (stacking fault, void, or interstitial) while green, yellow, and red atoms are grain boundary plane, triple junction, and vertex points, respectively. (c) A single grain surrounded by the various grain boundary network atom types.

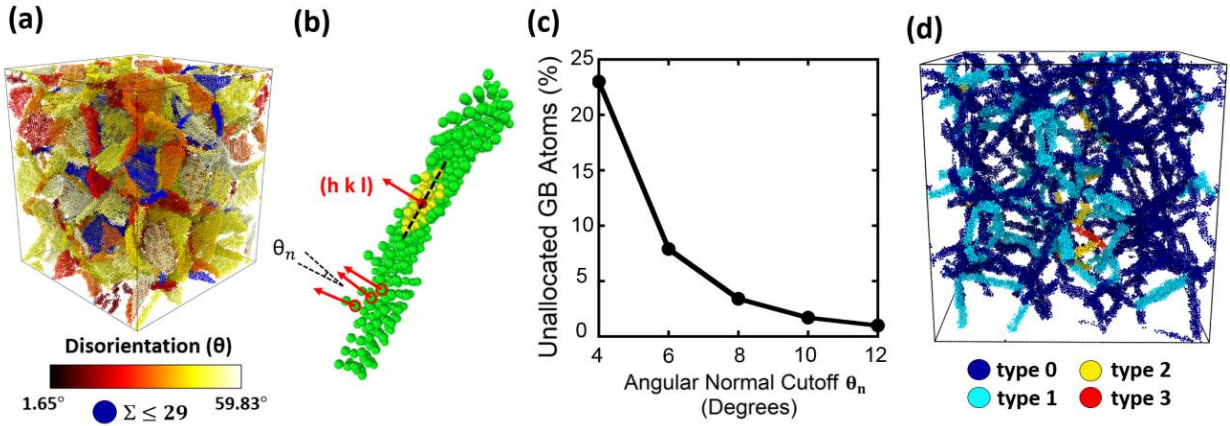

**Figure 3.** **(a) Grain boundary plane atoms colored according to disorientation angle. Special boundaries are highlighted in blue. (b) Example of a cluster of grain boundary plane atoms where a specific atom (red) and 50 of its neighbors (yellow) were used to calculate a local plane normal. This process is repeated until all grain boundary interface atoms are assigned a normal vector. Neighboring normal vectors must deviate by less than a specified angular cutoff value in order to be considered part of the same boundary plane. (c) The effect of varying the angular cutoff on the number of grain boundary interface atoms which are allocated to planar sections. (d) The triple junction network colored according to junction type.**

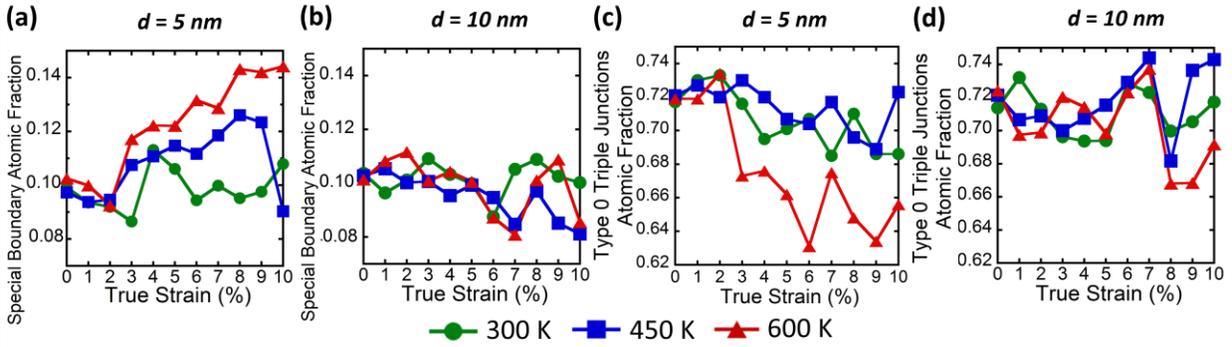

**Figure 4.** The evolution of special boundary fraction for (a) *d* = 5 nm and (b) *d* = 10 nm monotonic tension tests at all testing temperatures. (c) Type 0 triple junction analysis for *d* = 5 nm sample, showing evolution which was inversely proportional to the special boundary fraction evolution. (d) Type 0 triple junction fraction for the *d* = 10 nm grain size exhibits fluctuations, but no discernable upward or downward trend.

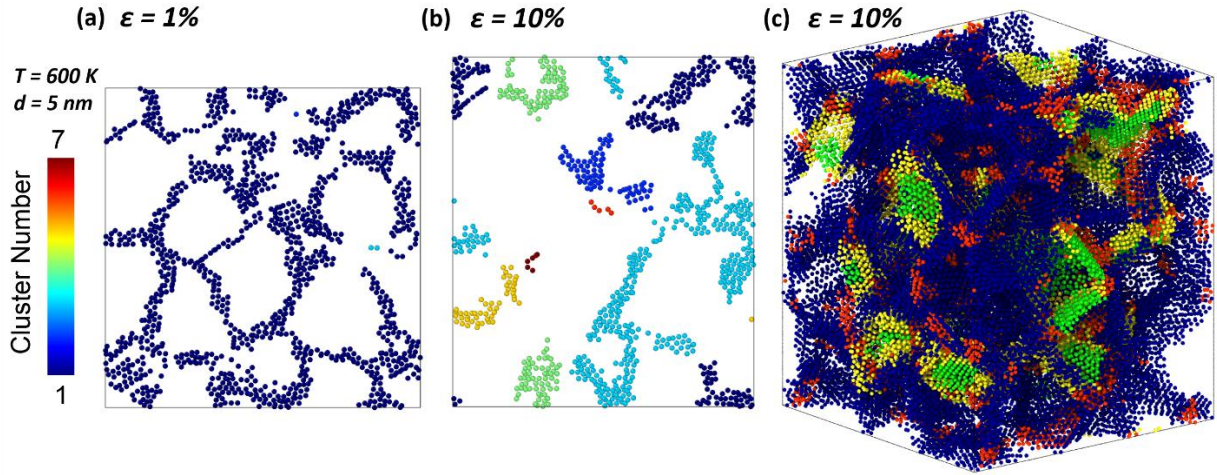

**Figure 5.** Two-dimension cluster analysis of the $d$ = 5 nm grain boundary network. For both (a) and (b) only those atoms which are not special nor are part of a special junction type are analyzed, with the colors associated with different boundary clusters. After being pulled to 10% strain, a clear breakup in the random network is observed. (c) The same sample with special boundaries (green), triple junctions (yellow) and vertex points (red) which were removed from the cluster analysis. The remaining atoms (blue) make up a single cluster which maintains connectivity throughout the sample.

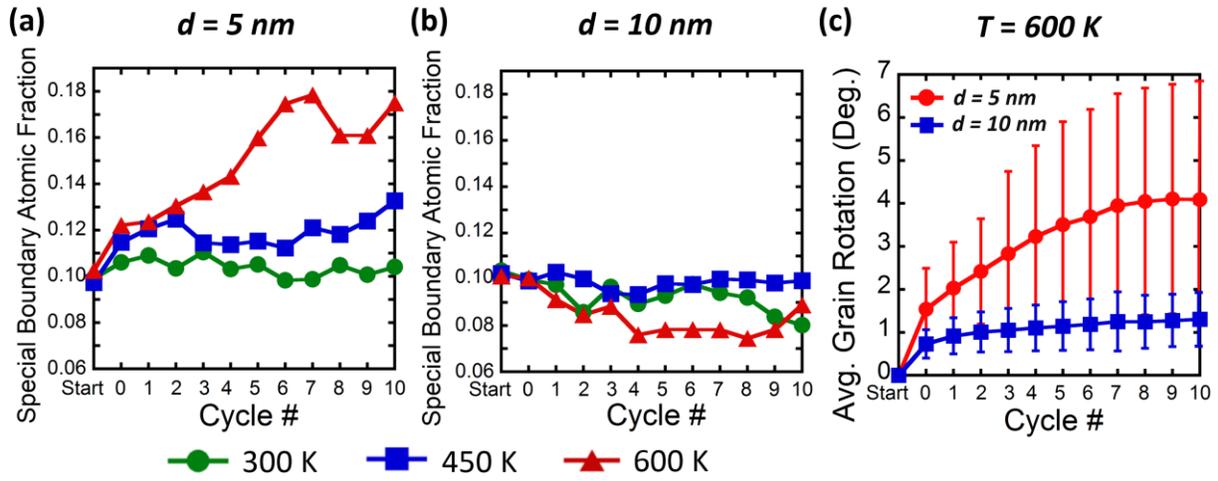

**Figure 6.** Special boundary fraction of mechanically cycled samples for (a) *d* = 5 nm and (b) *d* = 10 nm for all testing temperatures. (c) Average grain rotation, measured as the disorientation from a grain's starting configuration, for 5 nm and 10 nm average grain sizes cyclically loaded at 600 K.

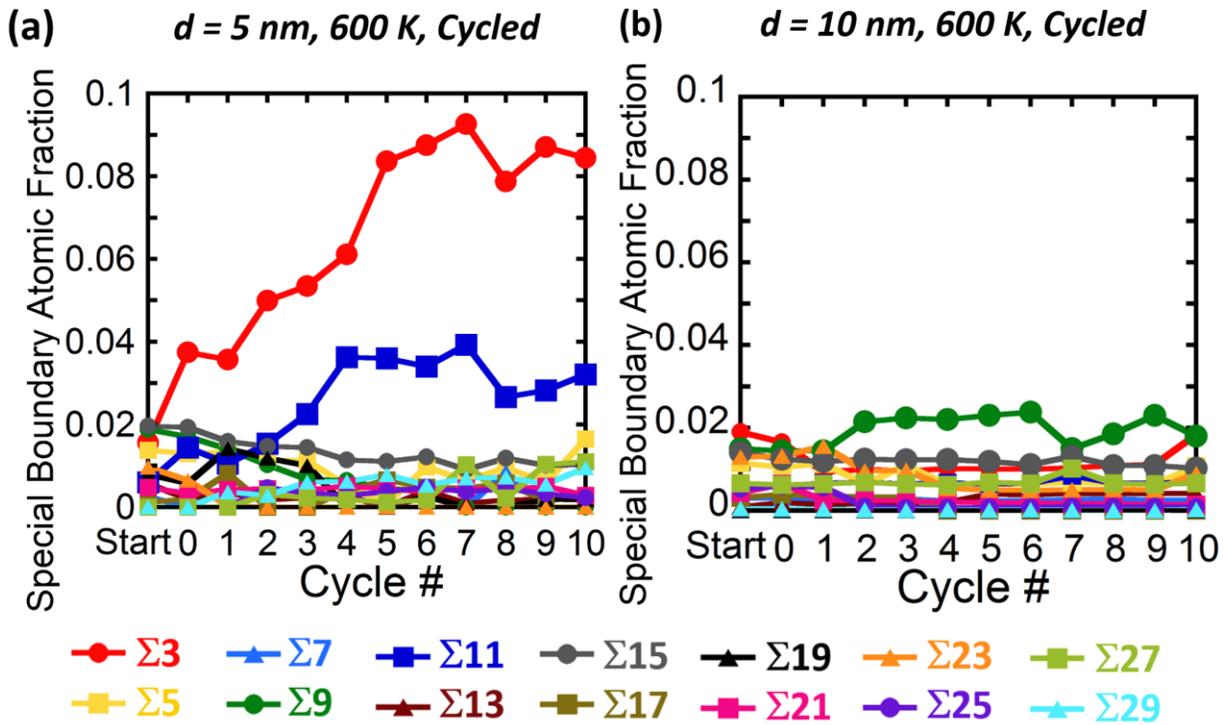

**Figure 7.** Special boundary fraction for the (a) *d* = 5 nm at 600 K sample shows an increase in Σ3 and Σ5 boundary content. (b) The *d* = 10 nm samples experience very little evolution of the special boundary fraction during cycling.

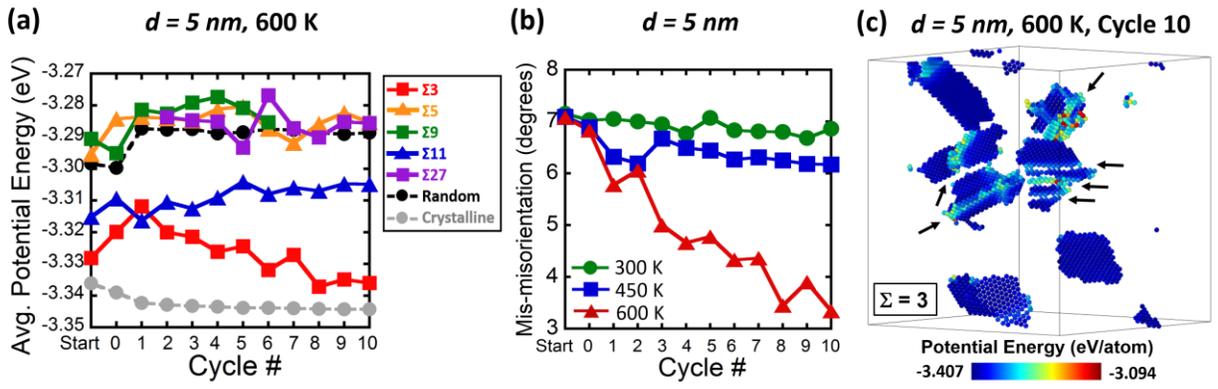

**Figure 8. (a) Average potential energies of select special boundary atoms, random boundary atoms, and crystalline atoms during cycling. (b) Mis-misorientation of Σ3 boundaries as a function of cycle number. (c) Σ3 boundaries at Cycle 10, with black arrows indicating facet steps which allowed coherent segments to form.**

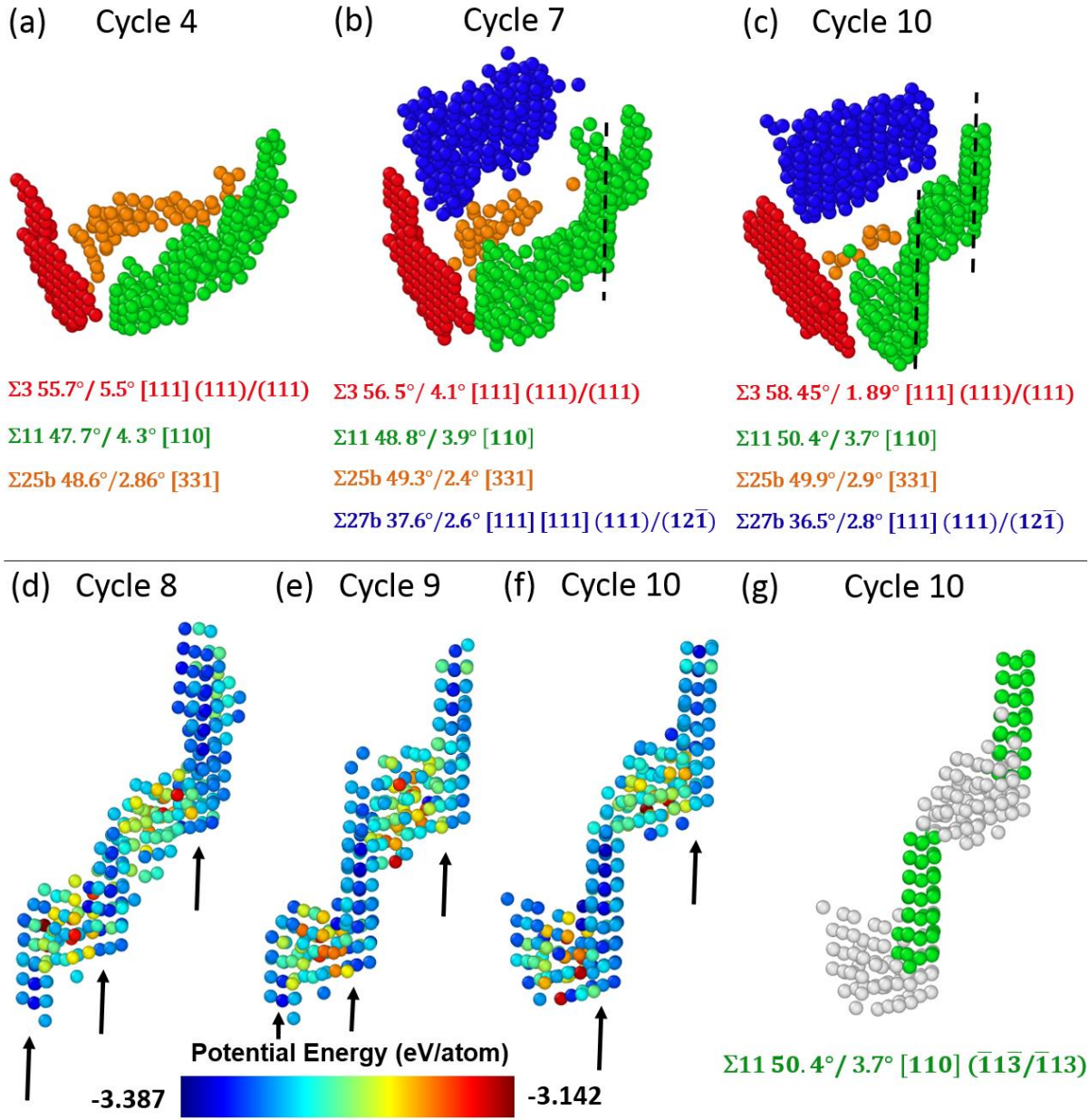

**Figure 9.** (a)-(c) The formation and evolution of selected special boundaries, with their Σ type, disorientation, mis-misorientation, and boundary plane information listed below each image where applicable, during mechanical cycling of a *d* = 5 nm sample at 600 K. (d)-(f) Cyclic loading drives a reduction in the energy the Σ11 boundary through grain boundary faceting. (e) The individual facet planes along the Σ11 boundary with low energy are identified as having a {113} boundary plane normal.

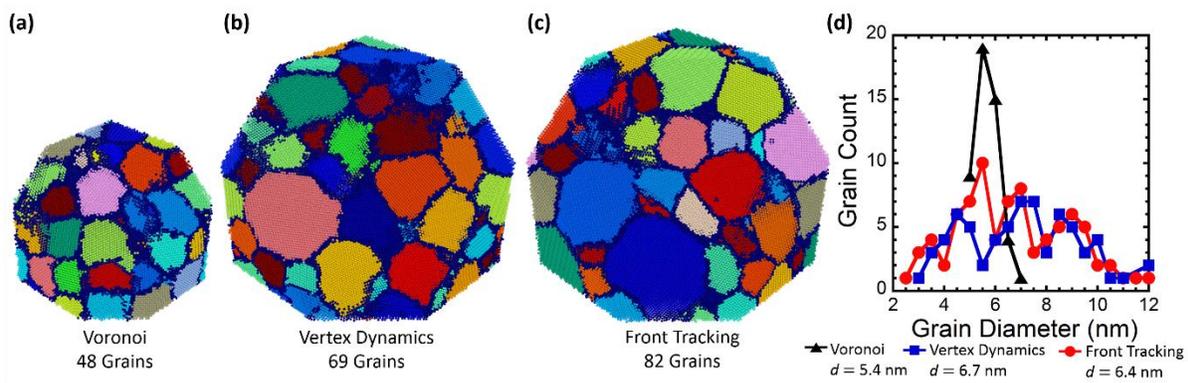

**Figure 10. (a)-(c) Visual comparison of the three different starting configurations post annealing at 600 K with grains colored according to GTA identification. (d) The grain size distribution for all three samples before cycling was performed showing a much narrower distribution for the Voronoi sample. Calculated average grain size for each sample are also listed below the figure.**

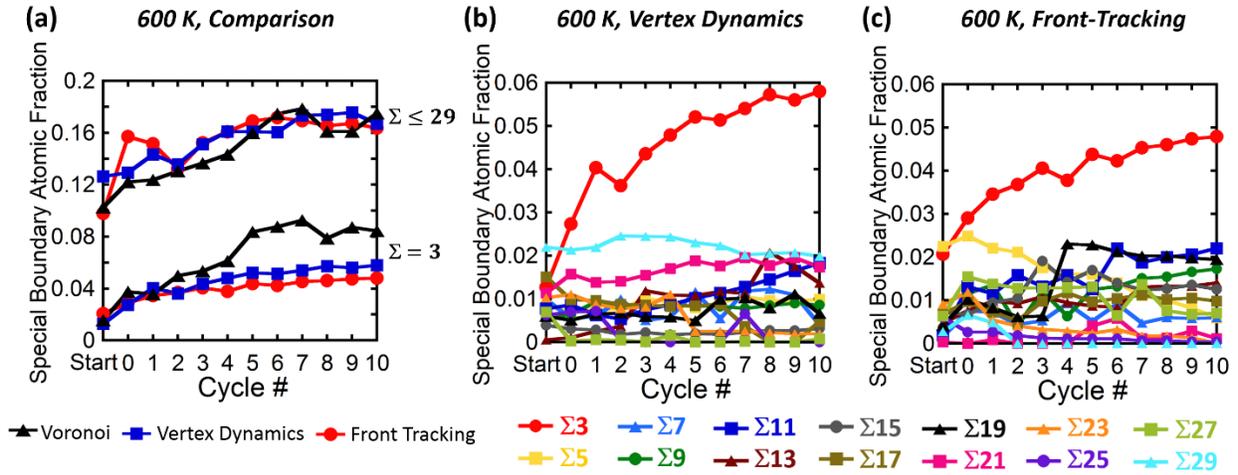

**Figure 11.** (a) **The upper curves show the special boundary fraction evolution for all boundaries with $\Sigma \leq 29$ as a function of cycle number. The lower curves compare the difference in $\Sigma 3$ evolution for all three starting configurations. Detailed evolution of special boundary fraction for all special $\Sigma$ types is shown as a function of cycle number for the (b) Vertex Dynamics and (c) Front-Tracking samples, respectively.**

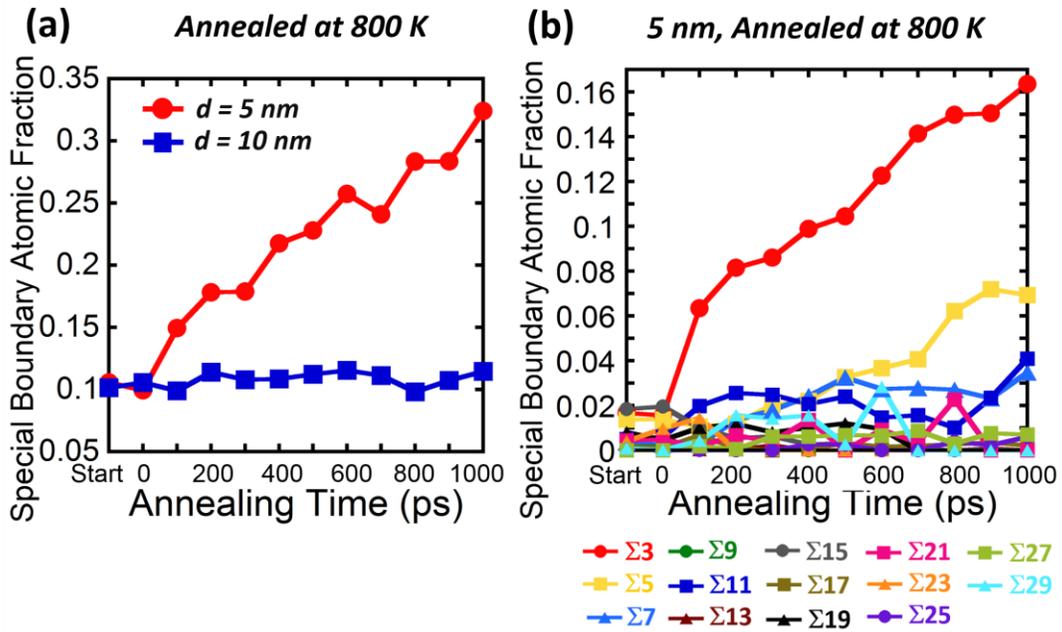

**Figure 12. (a)** Special boundary fraction for both grain sizes as a function of annealing time at 800 K. **(b)** A detailed breakdown of each special Σ type for the d = 5 nm sample, showing that Σ3 boundaries exhibit the fastest increase while Σ5, Σ7, and Σ11 interfaces also become more common.

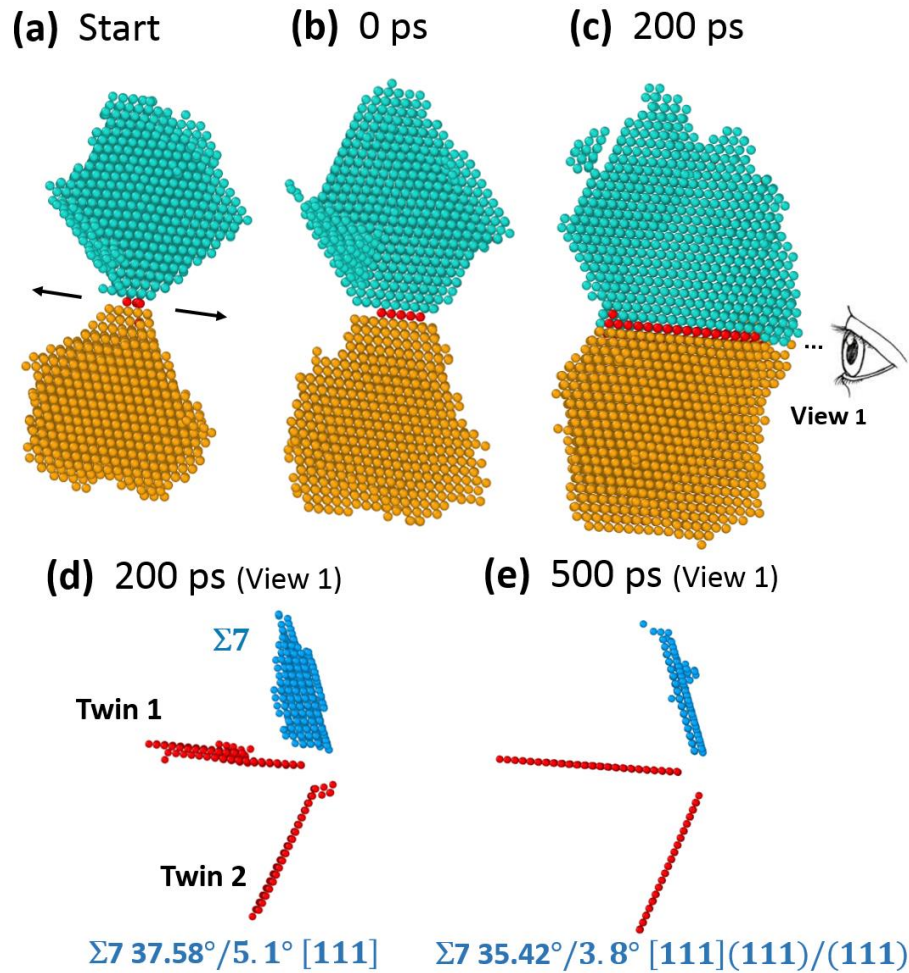

**Figure 13.** (a)-(c) Curvature-driven grain boundary migration leads to Σ3 boundary lengthening in the *d* = 5 nm sample during annealing. Additional evidence of grain boundary facet removal is shown in (d) and (f), where a well-defined (111) plane forms along the Σ7 boundary after 500 ps of annealing.